\documentclass[aps,prd,notitlepage,nofootinbib,onecolumn,groupedaddress,showkeys,floatfix]{revtex4-2}

\usepackage{graphicx}
\usepackage{amssymb}
\usepackage{amsmath}
\usepackage{bm}

\begin{document}


\title{Collapse of wave functions in Schr\"odinger's wave mechanics}





\author{Rainer Dick}
\email{rainer.dick@usask.ca}
\affiliation{Department of Physics and Engineering Physics, 
University of Saskatchewan, 116 Science Place, Saskatoon, Canada SK S7N 5E2}



\begin{abstract}
We show that inelastic scattering
  leads to a collapse of the wave function within standard evolution through
  the Schr\"odinger equation, whereas elastic scattering will not collapse
  the wave function. Specifically, we find that 
  the initial width of the emerging wave function in inelastic scattering
  is primarily determined by the size of the participating scattering center,
  but not by the width of the incoming wave function. 
  This implies that dynamical collapse of the wave function
  through inelastic scattering, together with energy quantization in bound quantum systems,
  can explain the emergence of particle-like signals without the need to invoke
  the Born rule.
\end{abstract}


\maketitle



The Born interpretation of the wave function $\psi(\bm{x},t)$ of a nonrelativistic
particle asserts that the probability to find the particle in a volume $V$
at time $t$ is
\begin{equation}
P_V(t)=\int_V\!d^3\bm{x}\,|\psi(\bm{x},t)|^2.
\end{equation}
Furthermore, the spatial extension of the signal from an individual particle
will be determined by detector resolution, but not by the width of the
probability density $|\psi(\bm{x},t)|^2$.
This agrees with observations in single-particle diffraction
experiments \cite{rf:merli,rf:tonomura,rf:bach}. Interference effects generate
macroscopic separations between different maxima of $|\psi(\bm{x},t)|^2$
and sampling the signals from many particles produces the interference patterns.
However, every single particle signal is still pointlike in the sense of being
determined by detector resolution.

These observations encode the ``measurement problem'' of quantum mechanics
in a nutshell. The continuous evolution of the wave function according to the
Schr\"odinger equation can produce mesoscopic or
macroscopic distributions $|\psi(\bm{x},t)|^2$,
and the confirmation of interference patterns from summation of many single-particle
signals confirms that evolution of the single-particle wave function probes
\textit{all the relevant scattering centers or paths on the way to the detector},
and yet the final individual single-particle
signal in the detector is pointlike and may involve interaction with only
a single atom-scale scattering center (e.g.~on a fluorescent screen), or a
sequence of pointlike interactions with consecutive single scattering centers
(e.g.~in a cloud chamber or bubble chamber), thus producing a ``classical''
particle track.

Motion from an electron gun to an electron detector produces a wave function
that corresponds to a coherent
superposition of scattering amplitudes from several scattering centers
(in order $e^2$ of the sum of amplitudes, i.e.~the particle did not
interact with one scattering center after the other), but the
act of detection seems to break this coherent evolution and suddenly the particle
interacts with only one scattering center in order $e^2$ of the final scattering
amplitude. The very act of ``observing'' or ``recording''
the particle with a detector seems to break the continuous wave-like evolution
of the wave function,
although the particle had simultaneously probed several scattering centers
over many atomic scales, or even over macroscopic distances, before the detection.

This led to the speculations about a special role of observers in the
old Copenhagen interpretation of quantum mechancis:
The combination of scattering \textit{and} observation does not 
register any pre-existing orbit of the particle but \textit{creates}
that orbit \cite{rf:wh1,rf:bohr,rf:bohr2,rf:wh2,rf:jvn}.
Heisenberg took a softer stance later in life and argued that localization
of the particle during the measurement process may also be a consequence
of complexity of interactions of the particle during the detection,
see Heisenberg's remarks in \cite{rf:wh3}, in particular p.~23.

The indeterminacy of particle properties is well documented in spin experiments
and widely accepted textbook knowledge in quantum mechanics.
However, any special role of detection (or observation or measurement or recording)
of a particle in localizing the particle, due to the presence or the actions
of a conscious observer, appears far fetched and did not
become generally accepted in the physics community.

The assumption that localization happens due to complexity of
interactions during the measurement process appears much more agreeable
and contributed to the motivation for the decoherence program, which
made important contributions to our understanding of the quantum-to-classical
transition and the emergence of classically preferred states
\cite{rf:zeh1,rf:zurek,rf:zeh4,rf:phz,rf:schlosshauer}.
However, complexity of interactions cannot solve the measurement problem
as such. The interaction of an electron with
the fluorescent screen in a low energy electron diffraction (LEED) experiment
is no more complex than its coherent interaction
with the surface atoms in the sample. Both kinds of interactions are described
in order $e^2$ through first-order scattering amplitudes of an electron interacting
with a medium, and there is no physical rationale for arguing that scattering
off the fluorescent screen should be more complex than scattering off the sample.
Indeed, the designation ``measurement problem'' is a misnomer that reflects
the fact that particle-like behavior was perceived to be linked to observations
in the early development of quantum mechanics. However, this is clearly not the case.
Once we trigger the electron gun in an LEED device,
the electrons will scatter coherently off the sample and produce pointlike flashes
on the screen, even when we do not watch nor turn on a CCD camera for recording.

It is tempting to infer from observation that the apparent break in wave function
evolution is a consequence of different behavior of the wave function
under elastic and inelastic scattering.
Generation of interference effects from
simultaneous contributions of several scattering centers or different paths requires
elastic scattering according to the Laue conditions, while registration in a detector
requires energy deposition through inelastic scattering.
Inelastic scattering in a medium can generate pointlike
signals, e.g.~on a scintillation screen, or tracks in a liquid medium, e.g.~in cloud
chambers or bubble chambers.

Particle detection through inelastic scattering implies that we deliberately
generate situations
where inelastic scattering is likely to occur. However, the premise of science to
identify laws of nature that desribe the universe independently from any
conscious measurement or observation, implies that a particle will behave in exactly
the same way irrespective of whether a particular medium is used for particle detection
or not. We would posit that most scientists
believe that a charged particle in a supersaturated vapor
will generate a track in every corner of the universe, irrespective of how far away from
the next life form or civilization. The question is then not whether observation or
measurement changes smooth evolution according to the Schr\"odinger equation, but
whether the Schr\"odinger equation itself can explain the emergence of particle
tracks.

Mott \cite{rf:mott} and Heisenberg \cite{rf:wh2} had pointed out that forward domination
of Coulomb scattering can naturally extend a particle track where a track has been
started. However, as Schonfeld correctly noted, the key question from the point of view
of the measurement problem is the formation of the beginning of the track, the first droplet
\cite{rf:jfs2}. Based on properties of polarized molecules in supersaturated vapors,
Schonfeld argues that Schr\"odinger evolution of the wave function
should be able to explain the emergence of the track through a ``consumption'' or ``draining''
of the initial spherical wave through the scattered wave function.

However, we would submit that every medium that can generate repeated inelastic particle
scattering, should generate a particle track, and this should not critically depend on
properties of the scattering centers.
We therefore model the medium as a set of scattering centers
at locations $\bm{a}_I$, where $I$ enumerates the scattering centers. Furthermore,
we assume fixed location of the scattering centers, as in a solid matrix. We are not
aiming for realistic modeling of liquid-medium detectors. We are rather interested in
a proof-of-principle investigation whether inelastic scattering 
can trigger wave function collapse within the scope
of Schr\"odinger's wave equation.

The general form of the model Hamiltonian for the particle-plus-medium system is then
\begin{eqnarray}\nonumber
  H&=&H_0+V
  \\ \label{eq:model1}
  &=&\frac{\bm{p}^2}{2m}+\sum_I\left(\frac{\bm{P}_I^2}{2M}+U(\bm{y}_I-\bm{a}_I)\right)
   +\sum_IV(\bm{x}-\bm{y}_I),
\end{eqnarray}
where $\bm{p}$ and $m$ are momentum and mass of the particle, and $\bm{P}_I$ and $M$
refer to momenta and mass of the scattering centers.
The internal potentials $U(\bm{y}_I-\bm{a}_I)$ generate the spectrum of the scattering
centers and $V(\bm{x}-\bm{y}_I)$ is the short-range
scattering potential of the particle with
the $I$-th scattering center. This would generically be a screened Coulomb potential
in realistic models.

We also assume that the scattering centers before scattering
are in their ground state with wave function $\phi_0(\bm{y}_I-\bm{a}_I)$. For example,
inelastic scattering of low-energy $\alpha$ particles (with energies of order
of a few 100 keV to a few MeV) in a solid material
primarily excites vibrations of the lattice
of nuclei, because the electronic form factors for momentum
transfers $\Delta p=\hbar\Delta k\gtrsim 100\,\mathrm{keV}/c$
satisfy $F(\Delta k)\ll 1$. Therefore, assuming harmonic oscillators
of frequency $\omega$, $U(\bm{y}_I-\bm{a}_I)=M\omega^2(\bm{y}_I-\bm{a}_I)^2/2$,
is not completely unrealistic as a first approximation for the scattering centers
in this case.

The Hamiltonian would evolve an initial
state $|\Psi(t')\rangle=\exp(-\,\mathrm{i}H_0t'/\hbar)|\Psi_i\rangle$
into the state $|\Psi(t)\rangle=\exp(-\,\mathrm{i}H_0t/\hbar)|\Psi_f\rangle$.
Here both the initial and the final state of the particle-plus-medium system are
expressed in terms of fiducial states
$|\Psi_i\rangle$ and $|\Psi_f\rangle$ at time $0$, where the fiducial states
are related to the actual system state at times $t'$ and $t$ through free time
evolution, i.e.~we describe the state before and after inelastic scattering
through asymptotic free states. This complies with the assumption of finite
range of the scattering potential $V$.

Canonical time evolution of $|\Psi(t)\rangle$ with the Schr\"odinger equation
implies
\begin{eqnarray}\nonumber
  |\Psi_f\rangle&=&\exp(\mathrm{i}H_0t/\hbar)\exp[-\,\mathrm{i}H(t-t')/\hbar]
  \exp(-\,\mathrm{i}H_0t'/\hbar)|\Psi_i\rangle
  \\ \label{eq:Psifi}
  &=&\mathrm{T}\exp[-\,\mathrm{i}\int_{t'}^t\!d\tau\,V_D(\tau)/\hbar]|\Psi_i\rangle,
\end{eqnarray}
which is equivalent to the scattering matrix with the Dirac picture Hamiltonian
\begin{equation}
V_D(t)=\exp(\mathrm{i}H_0t/\hbar)V\exp(-\,\mathrm{i}H_0t/\hbar).
\end{equation}
The connection with the scattering matrix
\begin{equation}
  S_{nm}(t,t')=\langle\Psi^{(0)}_n|
  \mathrm{T}\exp[-\,\mathrm{i}\int_{t'}^t\!d\tau\,V_D(\tau)/\hbar]|\Psi^{(0)}_m\rangle
\end{equation}
becomes explicit if we expand $|\Psi_f\rangle$ and $|\Psi_i\rangle$
in terms of eigenstates $|\Psi^{(0)}_n\rangle$ of $H_0$,
\begin{equation}
|\Psi_f\rangle=\sum_{n,m}|\Psi^{(0)}_n\rangle S_{nm}(t,t')\langle\Psi^{(0)}_m|\Psi_i\rangle.
\end{equation}

The scattering matrix contains all possible scattering channels of the particle-plus-medium
system, and we focus on first-order scattering off the scattering potential $V$.
 First-order scattering already includes the possibility of elastic scattering from
all the scattering centers. This maps the particle momentum $\bm{p}_i\to\bm{p}_f$
with $|\bm{p}_f|=|\bm{p}_i|$, and all coherently illuminated scattering centers contribute
to the amplitude that takes $\bm{p}_i$ to $\bm{p}_f$. Inelastic scattering, on the other
hand, leaves an imprint on the medium in exciting a scattering center in first
order.
The magnitudes of the scattering matrix elements determine which transitions are more likely
to happen. Surfaces tested in LEED devices yield electron scattering matrices that are
dominated by elastic terms, while any of the subdominant inelastic terms decrease the
signal-to-noise ratio. Electron detection devices, on the other hand, should yield
scattering matrices that are dominated by inelastic terms to detect electrons through energy
transfer to individual scattering centers.

We assume that inelastic scattering excites the scattering center at location $\bm{a}_I$
from the ground state with wave function $\phi_0(\bm{y}_I-\bm{a}_I)$ and energy $E_0$
into an excited state with wave function $\phi_1(\bm{y}_I-\bm{a}_I)$ and energy $E_1$.
The other scattering centers do not contribute in leading order of the scattering
potential, and we can represent the state as a
two-particle wave function $\langle\bm{x},\bm{y}_I|\Psi(t)\rangle$.
The short range of the scattering potential then
implies that we can write the two-particle state before and after the scattering
in the form
\begin{eqnarray}
  |\Psi(t')\rangle&=&\exp(-\,\mathrm{i}H_0t'/\hbar)|\psi_i\rangle\otimes |\phi_0\rangle,
  \\
  |\Psi(t)\rangle&=&\exp(-\,\mathrm{i}H_0t/\hbar)|\psi_f\rangle\otimes |\phi_f\rangle.
\end{eqnarray}

 Free time evolution generates dispersion of wave packets with a time scale
 $\tau=2m\Delta x^2/\hbar$.
 However, the free dispersion time scale is much larger than
particle travel times in lab experiments
with electrons from millimetre aperture electron guns
or neutrons from centimetre aperture neutron pipes, see, e.g.~\cite{rf:rdqm3ed}.
Our condition for emergence of a pointlike signal from wave function collapse is
therefore that the width of $\langle\bm{x}|\psi_f\rangle$ is primarily determined
by the size of the scattering centers, but not by the width
of $\langle\bm{x}|\psi_i\rangle$. 

Projection of the time evolution (\ref{eq:Psifi}) into the single-particle sector yields
in leading order of the scattering potential
\begin{eqnarray}\nonumber
  |\psi_f\rangle&=&-\,\frac{\mathrm{i}}{\hbar}
  \int_{t'}^t\!d\tau\,\exp(\mathrm{i}\omega_{10}\tau)
  \int\!d^3\bm{y}\,
  \exp\!\left(\mathrm{i}\frac{\bm{p}^2\tau}{2m\hbar}\right) V(\bm{x}-\bm{y})
  \\  \label{eq:Sfi1}
  &&\times
  \phi^+_1(\bm{y}-\bm{a}_I)\phi_0(\bm{y}-\bm{a}_I)
  \exp\!\left(-\,\mathrm{i}\frac{\bm{p}^2\tau}{2m\hbar}\right)|\psi_i\rangle,
\end{eqnarray}
where $\bm{p}$ and $\bm{x}$ are the operators for the scattered particle.
The transition frequency $\omega_{10}$ corresponds to the energy transfer from
the particle to the scattering center,
\begin{equation}
  \omega_{10}=(E_1-E_0)/\hbar.
\end{equation}

Evaluation of Eq.~(\ref{eq:Sfi1}) in the usual limits $t\to\infty$, $t'\to -\,\infty$
(which practically only means that $t-t'$ should be large compared to the travel
time of the particle through the range of the scattering potential),
yields the outgoing wave packet $\psi_f(\bm{x})$ in terms of the incoming wave
packet $\psi_i(\bm{x})$, both taken at fiducial time $0$,
\begin{eqnarray}\nonumber
  \psi_f(\bm{x})&=&\frac{m}{(2\pi)^3\mathrm{i}\hbar^2}\int\!d^3\bm{y}\int\!d^3\bm{z}
  \int\!d^3\bm{x}'\int\!d^3\bm{k}\,\exp[\mathrm{i}\bm{k}\cdot(\bm{x}-\bm{z})]
  \\ \nonumber
  &&\times
  \frac{\sin\!\left(\sqrt{k^2+(2m\omega_{10}/\hbar)}|\bm{z}-\bm{x}'|\right)}{
    \pi|\bm{z}-\bm{x}'|}\, V(\bm{z}-\bm{y})
  \\ \label{eq:psif1}
  &&\times\phi^+_1(\bm{y}-\bm{a}_I)\phi_0(\bm{y}-\bm{a}_I)\psi_i(\bm{x}').
\end{eqnarray}

The wave packet $\psi_f(\bm{x})$ is not yet normalized because it emerged from
the unitary time evolution (\ref{eq:Psifi}) through projection onto
the excited state $\phi_1(\bm{y}-\bm{a}_I)$ of the scattering center.
This yields only a conditional probability for this particular
inelastic scattering process
relative to the other elastic and inelastic scattering channels that are
possible in (\ref{eq:Psifi}). However, we can infer interesting information
on the width of inelastically scattered wave packets from Eq.~(\ref{eq:psif1}).

The Dirichlet kernel $\sin\!\left(\kappa|\bm{z}-\bm{x}'|\right)/(\pi|\bm{z}-\bm{x}'|)$
in Eq.~(\ref{eq:psif1}) provides an {\AA}ngstr\"om-scale approximation to
a $\delta$-function already for $k=0$ and for particle
masses $m\ge m_e=511\,\mathrm{keV}/c^2$
  if the excitation energy satisfies $\hbar\omega_{10}\gtrsim 1\,\mathrm{eV}$.
  Furthermore, the factor $\phi^+_1(\bm{y}-\bm{a}_I)\phi_0(\bm{y}-\bm{a}_I)$ will provide
  atom-scale resolution around the location $\bm{a}_I$ of the scattering center, and
  screened Coulomb potentials have {\AA}ngstr\"om-scale range in dense materials.
  All this indicates that the emerging wave packet from inelastic scattering should
  be located at the scattering center and have an initial width that is determined
  by the width of the scattering center, in perfect agreement with the operational
  definition of a particle-like signal. Indeed, approximating the {\AA}ngstr\"om-scale
  factors in Eq.~(\ref{eq:psif1}) with $\delta$-functions yields
  \begin{equation}\label{eq:collapse}
    \psi_f(\bm{x})\sim \phi^+_1(\bm{x}-\bm{a}_I)\phi_0(\bm{x}-\bm{a}_I)\psi_i(\bm{x}).
  \end{equation}
  The atomic transition factor $\phi^+_1(\bm{x}-\bm{a}_I)\phi_0(\bm{x}-\bm{a}_I)$
  will cut a piece of atom-scale resolution out of the wider wave function of
  the incoming particle, and this piece will be anchored around the scattering
  center.
  ``Pointlike'' position detection of subatomic particles
  \textit{per se} is therefore not a miracle in wave mechanics.

    We note that this derivation of wave function collapse works only for
    inelastic scattering. The scattered wave function for elastic scattering will
    always be a superposition of the 0th order term $\psi^{(0)}_f(\bm{x})\sim\psi_i(\bm{x})$
    and a first order term $\psi^{(1)}_f(\bm{x})$.
    The 0th order term implies that the elastically scattered wave packet should at least inherit
    the width of the incoming wave packet. Furthermore, the Dirichlet kernel for elastic
    scattering contains no energy transfer term $2m\Delta E/\hbar^2$ in the argument, and this
    implies that the Dirichlet kernel will not provide an atom-scale approximation to
    a $\delta$-function for small particle momenta. The scattering matrix tells us that
    wave function collapse will happen in inelastic scattering, but not in elastic scattering.
    
    However, wave function collapse in inelastic scattering by itself
    may not absolve us of the measurement problem in wave mechanics.
  The catch is that the mathematical evolution of the wave function
  through inelastic scattering might also allow for the superposition of two inelastic
  scattering
  events in locations $\bm{a}_1$ and $\bm{a}_2$. This would yield the
  superposition of a pointlike particle signal in location $\bm{a}_1$
  due to excitation of the scattering center $\bm{a}_1$, with another
  pointlike particle signal in location $\bm{a}_2$
  due to excitation of the scattering center $\bm{a}_2$.
  In the absence of any additional restrictions on permissible final states,
  coherent superposition of two inelastic scattering events appears to correspond to
  an allowed final state of the scattering matrix.
  With respect to the detector, this would generate a superposition of
  states proportional to
  $|\phi_{1,\bm{a}_1}\rangle|\phi_{0,\bm{a}_2}\rangle+\exp(\mathrm{i}\beta)
  |\phi_{0,\bm{a}_1}\rangle|\phi_{1,\bm{a}_2}\rangle$
  (up to normalization, and with
  an arbitrary relative phase $\beta$). Without the Born rule, the
  Schr\"odinger equation might allow for the formation of more than
  one particle track, or more than one point lit on a fluorescent screen.

  This is not a problem in epistemic interpretations of
  the wave function
  \cite{rf:ballentine,rf:rovelli1,rf:epi1,rf:epi2,rf:rovelli2,rf:epi3}
  because the Born rule is understood as a necessary prescription for
  translating an epistemic wave function into a position signal.

  It is also not a problem in Bohmian
  mechanics \cite{rf:bohm,rf:duerr,rf:maudlin,rf:goldstein} because the
  inclusion of a quantum potential and the
  addition of a guiding equation make the Born rule redundant.

  Notwithstanding, wave function collapse in inelastic scattering opens up a third
  line of attack on the measurement problem. We may be able to exclude the simultaneous
  excitation of different scattering centers through the observation
  that low-energy excitations in
  isolated bound quantum systems can only yield energy eigenstates.
  This excludes states of the form\footnote{We assume that we can neglect
    interactions between different scattering centers.}
  $|\phi_{1,\bm{a}_1}\rangle|\phi_{0,\bm{a}_2}\rangle+\exp(\mathrm{i}\beta)
  |\phi_{0,\bm{a}_1}\rangle|\phi_{1,\bm{a}_2}\rangle$ as final detector states
  and explains the emergence of a single pointlike position signal without
  invoking the Born rule.

\acknowledgments
We acknowledge support from the Natural Sciences and Engineering Research Council
of Canada.\\

\end{document}